\documentclass[aps,amsmath,amssymb,floatfix,superscriptaddress,twocolumn,pra]{revtex4-2}

\usepackage{graphicx}%
\usepackage{multirow}%
\usepackage{amsmath,amssymb,amsfonts}%
\usepackage[title]{appendix}%
\usepackage{xcolor}%
\usepackage{booktabs}%
\usepackage{algorithm}%
\usepackage{algorithmicx}%
\usepackage{algpseudocode}%
\usepackage{listings}%
\usepackage{xspace}
\usepackage{physics}
\usepackage{bm}
\usepackage{comment}
\usepackage{hyperref}

\newcommand{\psh}[2]{\ensuremath{\langle #1|#2\rangle}\xspace}

\newcommand{\bfr}{{\bf{r}}}

\newcommand{\bfw}{{\bf{w}}}

\newcommand{\bmtheta}{{\bm{\theta}}}

\begin{document}


\title{Quantum Simulation of Electronic Structure using Quantum Density Functional Theory on a  Programmable Integrated Photonic Circuit}

\newcommand{\Sapienza}{Dipartimento di Fisica - Sapienza Universit\`{a} di Roma, P.le Aldo Moro 5, 00185 Roma, Italy}
\newcommand{\ICGM}{ICGM, Universit\'e de Montpellier, CNRS, ENSCM, Montpellier, France}
\newcommand{\IFN}{Istituto di Fotonica e Nanotecnologie, Consiglio Nazionale delle Ricerche (IFN-CNR), Piazza Leonardo da Vinci, 32, 20133 Milano, Italy}
\newcommand{\UF}{Unitary Fund France, 1 Impasse du Palace, 37000 Tours, France}
\newcommand{\Ephos}{Ephos, Viale Decumano 34, 20157 Milano, Italy}
\newcommand{\Athens}{Department of Physics, School of Applied Mathematical and Physical Sciences, National Technical University of Athens, 9, Ir. Politechniou Str, 157 80 Athens, Greece}

\author{Giovanni Adriano} \thanks{These two authors contributed equally}\affiliation{\Sapienza}
\author{Akilan Rajamani}\thanks{These two authors contributed equally}\affiliation{\ICGM}
\author{Francesco Hoch}\affiliation{\Sapienza}
\author{Farrokh Labib}\affiliation{\UF}
\author{Nadja Z. Zaouri}\affiliation{\Athens}\affiliation{\Sapienza}
\author{Eugenio Caruccio}\affiliation{\Sapienza}
\author{Gonzalo Carvacho}\affiliation{\Sapienza}
\author{Nicol\`o Spagnolo}\affiliation{\Sapienza}
\author{Riccardo Albiero}\affiliation{\IFN}
\author{Giacomo Corrielli}\affiliation{\IFN}\affiliation{\Ephos}
\author{Francesco Ceccarelli}\affiliation{\IFN}
\author{Jean-Sebastien Filhol}\affiliation{\ICGM}
\author{Roberto Osellame}\affiliation{\IFN}\affiliation{\Ephos}
\author{Taira Giordani}\email{taira.giordani@uniroma1.it}\affiliation{\Sapienza}
\author{Fabio Sciarrino}\affiliation{\Sapienza}
\author{Nathan Shammah}\email{nathan@unitary.foundation}\affiliation{\UF}
\author{Bruno Senjean}\email{bruno.senjean@umontpellier.fr}\affiliation{\ICGM}

\begin{abstract}
Quantum simulations of molecular electronic structure hold the potential for computational advantage but are inherently limited by the linear scaling of qubit requirements with respect to system size. Quantum Density Functional Theory (Q-DFT) addresses this constraint by reformulating the interacting many-body problem in terms of a non-interacting Kohn--Sham reference system. Here, we present and experimentally validate a Q-DFT algorithm that requires only $\mathcal{O}(\log_2(N))$ qubits to simulate a system of $N$ orbitals, thereby reducing resource overhead. More precisely, we solve the Kohn--Sham problem for the lithium hydride (LiH) molecule using the ensemble extension of the variational quantum eigensolver paradigm that minimizes the energy of the system in the subspace spanned by the ground and first excited states. To execute this task, we encode the problem with qubits on a quantum photonic platform, utilizing an architecture that combines a deterministic quantum dot single-photon source with a programmable integrated photonic circuit. By evaluating the molecular ground- and first excited-state properties, this work establishes a pathway for simulating larger molecular systems, demonstrating the viability of integrating qubit-efficient algorithms with advanced reconfigurable photonic hardware for near-term quantum chemistry applications.
\end{abstract}




\maketitle

\section{Introduction}\label{sec1}



Understanding the electronic structure of molecules is the cornerstone of quantum chemistry, enabling breakthroughs in material design, catalysis, and energy storage.
Although quantum chemistry has extensively used approximations such as Born-Oppenheimer for electron nucleus coupling, the multiconfigurational expansion of the electronic wavefunction description still grows exponentially with system size ({\it i.e.}, the number of orbitals), thus limiting fully accurate simulations up to a few atoms.
Nevertheless, material modelling approaches using classical computing mainly favour the use of density functional theory (DFT) over computing the many-body wavefunction. DFT  uses the one-body electronic density
-- derived from a non-interacting system --
as its fundamental variable, with mathematical foundations established by Hohenberg, Kohn, and Sham in the 1960s~\cite{hktheo,KS}.
Since then, DFT has revolutionised the field by enabling simulations of much larger systems,  becoming the workhorse method for industrial applications such as heterogeneous catalysis~\cite{filhol2004,norskov2011} or Li-ion~\cite{CEDER1997161, filhol2020} and post Li batteries~\cite{filhol2020b} modeling. 

More recently, the second quantum revolution, 
enabled by the application of the laws of quantum mechanics to information theory, 
offered a transformative alternative by taking advantage of the inherent quantum nature of the electronic structure problem \cite{Cao2019}.
Despite the fact that technology is not mature enough yet to lead to a clear quantum advantage over classical computers,
several quantum algorithms~\cite{kitaev1995quantum,aspuru2005simulated,cao2019quantum,bauer2020quantum,head2021quantum,ayral2023quantum,bharti2022noisy,motta2022emerging,zhang2025quantum} and experimental demonstrations~\cite{kandala2017hardware,arute2020hartree,rice2021quantum} have already been implemented
and show promises toward potential future breakthroughs,
for instance in battery materials and technologies~\cite{rice2021quantum,hariharan2025modeling,kim2022fault,zini2023quantum,delgado2022simulating,fomichev2025fast,kunitsa2025quantum}. 

In this context, photonics offers a highly effective platform for quantum computing and simulation, enabling high-fidelity quantum state preparation, manipulation and measurement \cite{Flamini_2019, Hasnaoui_rev}. 
Recent technological advancements in photonic quantum technologies have enabled the fabrication of fully programmable integrated photonic circuits \cite{Wang2019, Pelucchi2021}. These highly reconfigurable platforms can implement various quantum algorithms, notably Variational Quantum Eigensolvers (VQE) for computing molecular energy states \cite{Peruzzo2014, Santagati2018, Kim_qudit_VQE}. 
In the context of photonic quantum simulators, starting from the first proof-of-concept demonstrations of photonic VQE on molecules such as H$_2$ \cite{Peruzzo2014}, the field has witnessed both advances to improve optimization efficiency by exploiting quantum algorithms, such as gradient calculations via the photonic quantum parameter shift rule \cite{Hoch2025a, facelli2024exactgradientslinearoptics, Cimini2024}, and significant hardware progress. This progress has culminated in highly programmable and multi-purpose platforms driven by integrated photonics, which can natively encode a variety of algorithms \cite{Arrazola2021ProgrammableNanophotonic,maring2024versatile, Hoch2025, monbroussou2025photonic}, including VQE \cite{maring2024versatile} and molecular simulations~\cite{Wang2017, Sparrow2018, Arrazola2021ProgrammableNanophotonic}.
In fact, photonic encoding naturally provides a direct approach to simulation of molecular vibrational spectra by exploiting its intrinsic bosonic nature~\cite{Huh2015}. 
Extending its applicability beyond these native implementations to address the inherent challenge of using a photonic (hence bosonic) device to solve a fermionic problem, a hybrid scheme has been proposed, mixing non-interacting boson dynamics and classical computational chemistry methods on small molecules such as lithium hydride (LiH)~\cite{shang2025boson}.

\begin{figure}
    \centering
\includegraphics[width=1\columnwidth]{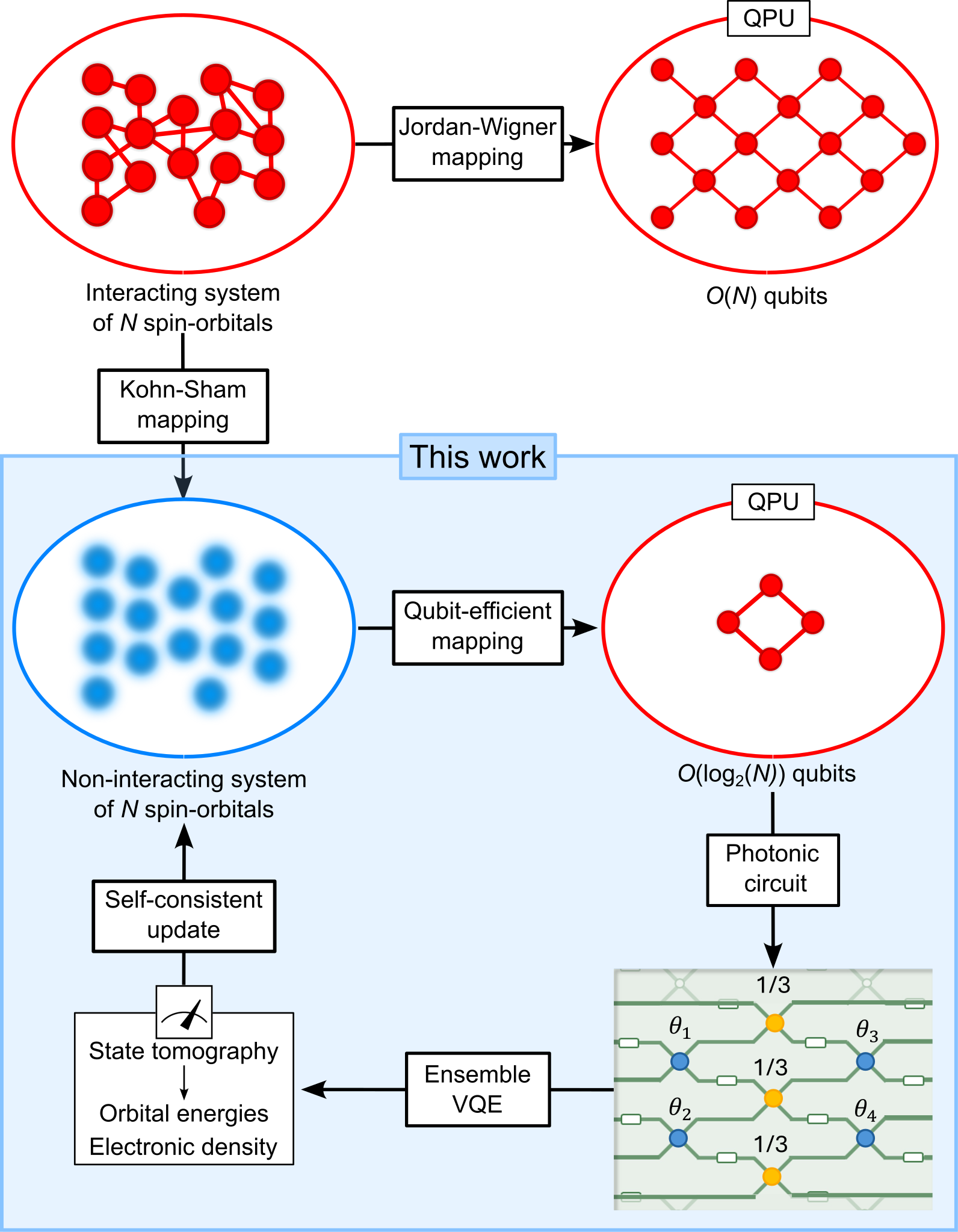}
    \caption{{\bf Illustration of the quantum implementation of Kohn--Sham DFT on a photonic circuit.}
    An interacting system of $N$ spin-orbitals is first mapped to 
    the auxiliary non-interacting Kohn--Sham system for which only $\log_2(N)$ qubits is needed rather than $N$ qubits.
    Using ensemble VQE and state tomography -- implemented on a photonic circuit described in Fig.~\ref{fig:8modechip} -- the ensemble energy and density of the Kohn--Sham system are estimated self-consistently to recover the ground-state energy of the physical (interacting) system. In this work, the self-consistent loop is only performed within the noiseless quantum simulation.}
    \label{fig:flowchart}
\end{figure}

In this work, we demonstrate a quantum simulation of the LiH molecule by implementing the quantum analogue of DFT (Q-DFT) on a programmable integrated photonic circuit, as illustrated in Fig.~\ref{fig:flowchart}. 
Originally developed in Ref.~\cite{senjean2023toward}, Q-DFT requires only $\mathcal{O}(\log_2(N))$ qubits to simulate a system of $N$ orbitals. 
This reduction in resource requirements enables quantum simulations of large systems, while maintaining an accuracy limited primarily by the approximate density functional employed~\cite{senjean2023toward}. 
Owing to this qubit-efficient mapping, only two qubits are required to simulate the LiH molecule.
To perform the optimization and determine an estimate of the electron density of the LiH molecule, we encode the corresponding Kohn--Sham DFT 
Hamiltonian into photonic qubits and perform a VQE ensemble optimization, a variant of the quantum eigensolver that minimizes the energy of the molecule in the subspace spanned by the ground and first excited states \cite{nakanishi2019subspace, yalouz2021state} (corresponding to the first two occupied orbitals within the Q-DFT framework). 
For this experimental task, we exploit a highly flexible quantum photonic architecture \cite{Hoch2025} that combines a multiplexed single-photon source, based on a deterministic quantum dot emitter \cite{Somaschi2016}, with a universal 8-port integrated photonic circuit fabricated via femtosecond laser writing \cite{Corrielli2021, Pentangelo2024}.
Alongside this proof-of-principle demonstration, we investigate the impact of finite-sampling and experimental noise on both the self-consistent convergence of the Q-DFT and the ensemble VQE optimization, while assessing the viability of scaling this approach and highlighting the synergy between qubit-efficient Q-DFT algorithms and integrated photonic platforms for molecular simulations.

\section{Results}\label{sec2}

\begin{figure}
    \centering
    \includegraphics[width=1\linewidth]{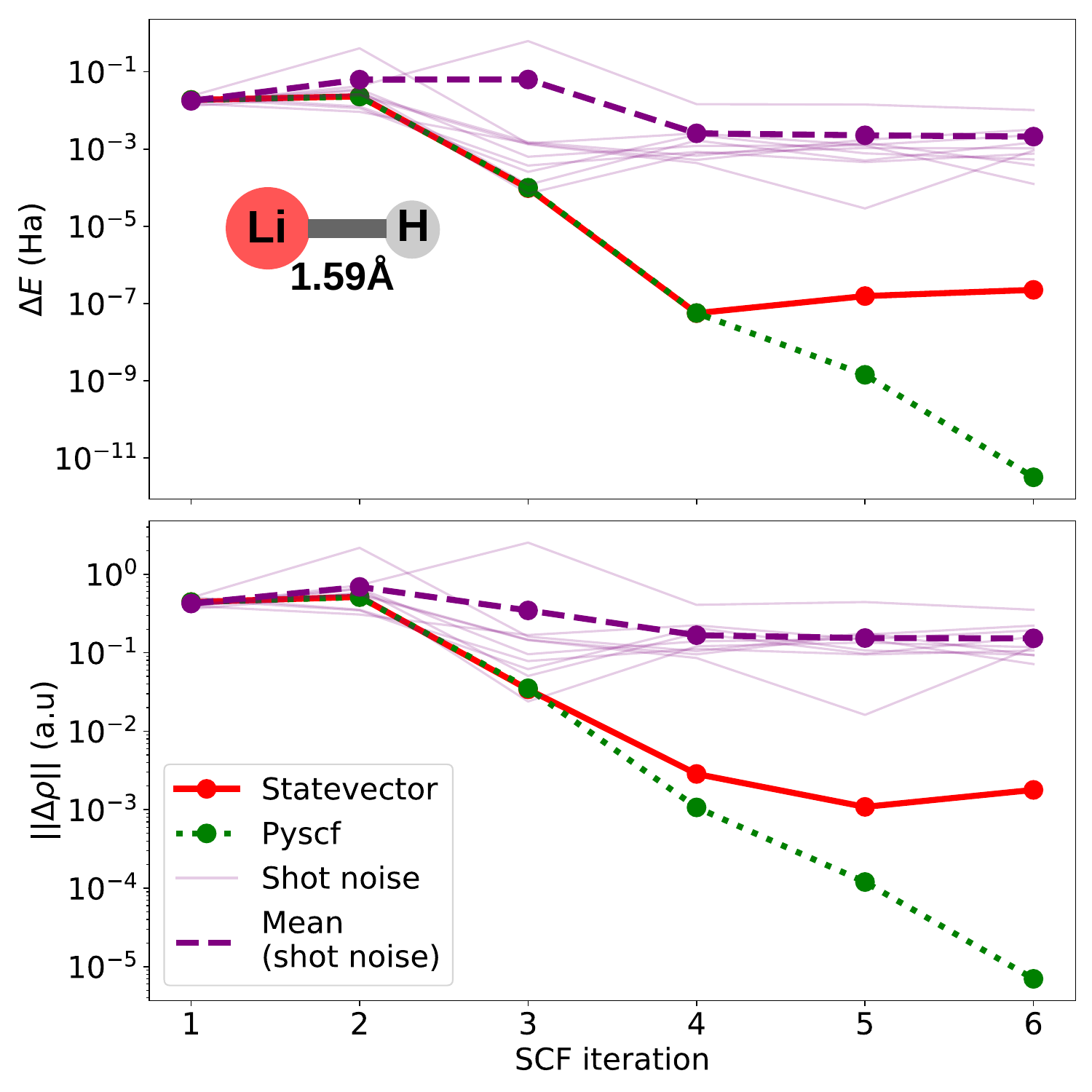}
    \caption{\textbf{Convergence of the self-consistent field (SCF) loop under exact and finite sampling conditions.}
    Errors in the ensemble energy (top panel) and in the electronic density (bottom panel) with respect to the converged result obtained with PySCF.
    The solid purple curves represent 10 simulations (initialized with random parameter values) using 10$^9$ shots, and the dashed line corresponds to their mean.}
    \label{fig:SCF_LiH}
\end{figure}

In this work, we focus on the quantum simulation of the lithium hydride molecule within Q-DFT,
using a noiseless simulation,
a noisy model simulation, and 
experimental simulation on a programmable integrated photonic circuit.

The electronic structure problem of the LiH molecule in the minimal basis set without any approximations requires 12 qubits using wavefunction theory, which remains challenging for current quantum hardware, real photonic devices included, due to platform-dependent limitations such as noise and scalability.
Hence, we move to the description of the LiH molecule using quantum density functional theory,
keeping only the 1s, 2s, and 2p$_z$ orbitals of the Lithium (the 2p$_x$ and 2p$_y$ do not directly
participate due to symmetry),
and the 1s of the hydrogen,
such that
only $\log_2(4) = 2$
qubits are needed to compute its ground-state energy
at the Kohn--Sham (KS) DFT level.
Note that removing the 2p$_x$ and 2p$_y$ orbitals would still require 8 qubits using, for instance, the Jordan--Wigner mapping on the many-body problem.
Since LiH has four electrons, they
occupy two KS orbitals, representing the ground and first-excited states
of the KS Hamiltonian that we prepare using the ensemble extension of the variational quantum eigensolver~\cite{nakanishi2019subspace,yalouz2021state,yalouz2022analytical,illesova2025transformation,rajamani2025equi}.
To comply with the circuit-depth constraints of the available photonic architecture for the implementation of the ensemble VQE required at each step of the Q-DFT, a single layer of the $R_Y$--C-NOT hardware-efficient ansatz is used, containing a single post-selected C-NOT gate operation (see Sec.~\ref{sec:methods} for more details on the method).

\subsection{Noiseless quantum simulation of LiH}

Before performing any experiments or noisy-model simulations, we emulate our quantum algorithm using state-vector simulations with the open-source software \texttt{qlass}~\cite{Labib2026}. 
Specifically, the software simulates the ensemble VQE circuit mapped onto the photonic encoding, paired with the classical COBYLA optimizer. 
 At each iteration, the ensemble energy is minimized using the simulated VQE, and the resulting electronic density is used to construct a new Hartree-exchange-correlation potential, {\it i.e.}, a new KS Hamiltonian that must then be solved again using ensemble VQE.
The noiseless Q-DFT simulation requires
six self-consistent iterations to reach convergence,
as shown in Fig.~\ref{fig:SCF_LiH}. The six identified Hamiltonians, expressed in the two-qubit Hilbert space as linear combinations of tensor products of single-qubit Pauli operators, are detailed in Appendix~\ref{app:pauli}.
The convergence criteria consist of an energy change below $10^{-5}$ Hartree and a change
in the root mean square deviation of the density below $10^{-4}$ between two consecutive iterations.
These iterations are compared with the reference PySCF results, which converge in 7 iterations.
As readily seen in Fig.~\ref{fig:SCF_LiH},
the state-vector simulation closely follows the reference energy for the first four iterations, and deviation starts to appear in the density at the fourth iteration.
As the self-consistent KS scheme is density-dependent, the subsequent iterations will be different by construction, deviating from PySCF.
However, the calculation still converges, with a final energy error below $10^{-6}$ and a density error around $10^{-3}$, compared to the reference.

Although the self-consistent cycle converges well in the absence of noise,
the energy difference between successive iterations is very small for a system of this size and is difficult to resolve in the presence, for example, of shot noise.
As a first step, we investigated the robustness of the protocol in such a noisy condition, a statistical uncertainty inherent to quantum hardware whenever energy expectation values are estimated from a finite ensemble of state copies prepared at each optimization stage. As shown in Fig.~\ref{fig:SCF_LiH} (purple curve), this finite-sampling noise directly impacts the convergence of the self-consistent cycle. More precisely, within the six SCF iterations fixed in our analysis, the simulations that include shot noise do not reach full convergence, even with a billion shots, settling instead into a plateau with an accuracy of approximately $10^{-3}$~Hartree for the energy and $10^{-1}$ for the electronic density. This behaviour stems from two main factors. First, SCF procedures are intrinsically sensitive to perturbations, making their convergence generally challenging. Second, due to the small scale of the molecular system investigated here, the KS Hamiltonians generated after the fourth iteration in the noiseless simulation become nearly identical (see also Appendix~\ref{app:pauli}). Consequently, the resulting energy differences between
successive steps are too small to be resolved within
the statistical accuracy imposed by shot noise, for sample size within the simulated regime. Further investigations are required to identify the scaling of the required sample size depending on the molecular system under analysis.
This problem could also be addressed in the near future by employing fault-tolerant quantum algorithms that are not subject to shot noise, and that can achieve arbitrarily low error rates by incorporating quantum error correction.

\subsection{Experimental photon-based quantum simulation of LiH}

\subsubsection{Experimental setup}

\begin{figure*}[t]
    \centering
    \includegraphics[width=\linewidth]{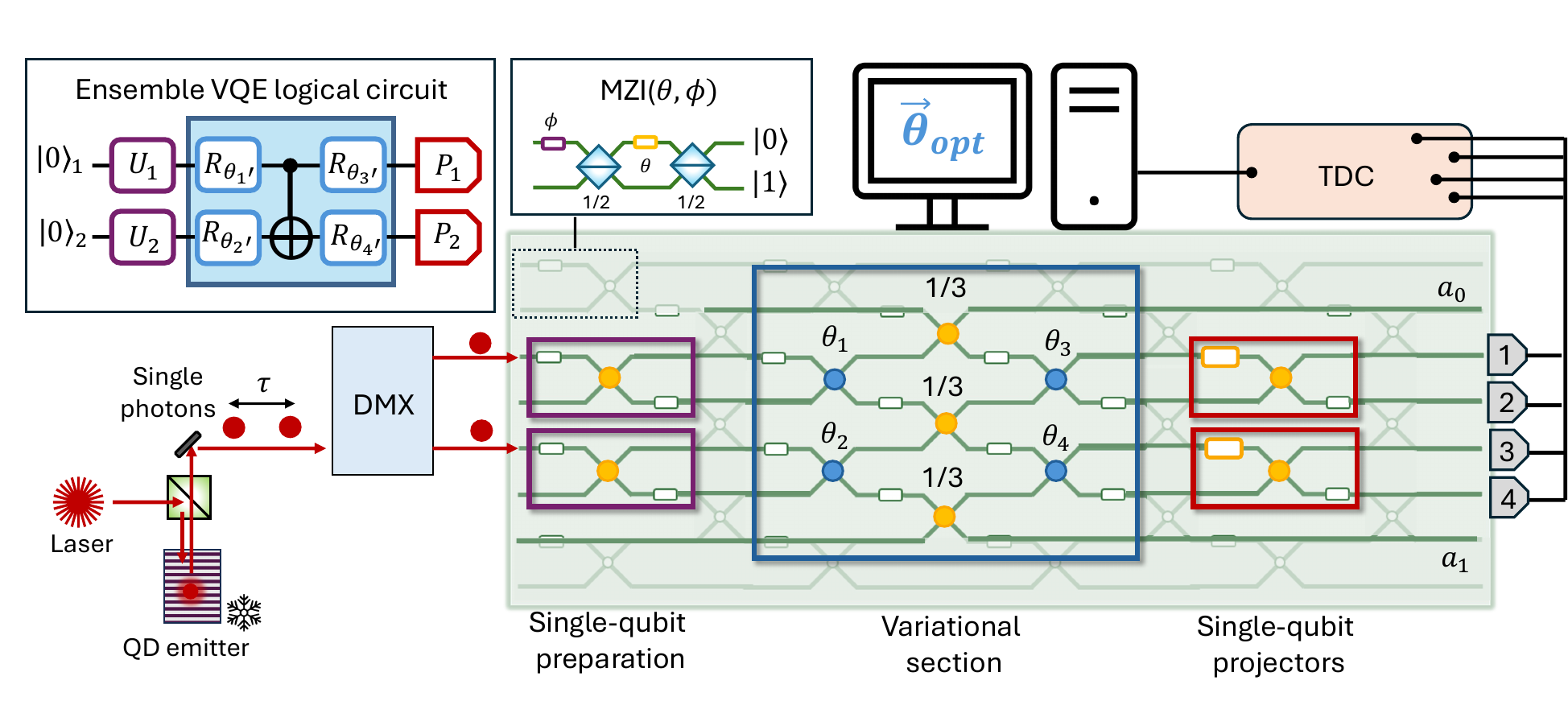}
    \caption{\textbf{Experimental setup.}
The photonic variational quantum eigensolver architecture employs a single-photon source based on a quantum dot (QD) emitter, which is excited by a pulsed laser in the resonance fluorescence regime. This is combined with a time-to-space demultiplexer (DMX) to sequentially route and temporally synchronize two consecutive single photons into the two input modes of the integrated photonic processor. The 8-mode photonic circuit follows a universal architecture, whose basic unit cell is a balanced Mach-Zehnder interferometer (MZI) detailed in the top-right inset. In this diagram, the circles represent tunable beam-splitters with equivalent reflectivities determined by the internal phase of the MZIs, while the rectangles represent external phase shifters. The colored components indicate the specific active elements utilized in this experiment. The top-left inset illustrates the $R_Y$--C-NOT logical circuit of the ensemble VQE, comprising a single-qubit preparation stage (purple), two layers of variational single-qubit rotations (blue) with an entangling C-NOT gate in between and single-qubit projectors on Pauli operators (red). The photonic architecture physically implements this scheme. In the single-qubit preparation stage, the initial MZIs can be configured to prepare the desired ansatz ensemble states, such as 
$b =  \{\ket{++}, \ket{--} \}$. In the main stage of the variational circuit, a photonic post-selected C-NOT gate (dark yellow MZI) is interposed between four real rotations $\vec{\theta}$ (blue MZI), which encode the four variational parameters $\vec{\theta}'$ as well as the MZI reflectivities of 1/2 required by the post-selected C-NOT. These four parameters ($\vec{\theta} = \theta_1, \dots, \theta_4$) are optimized iteratively by a classical computer to minimize the target cost function. Finally, a measurement stage (red rectangles) rotates the two qubits into the eigenbases of the Pauli operators. The coincidence events are then recorded by a time-to-digital converter (TDC) and then processed by the classical computer to evaluate the expectation values.
    }
    \label{fig:8modechip}
\end{figure*}

The experiment is performed on a quantum photonic platform~\cite{Hoch2025, rodari2024, monbroussou2025photonic}, summarized in Fig.~\ref{fig:8modechip}. 
A single-photon source based on a semiconductor quantum dot in the resonance fluorescence configuration is excited by a pulsed laser (see Methods for the typical values of single-photon purity and photons indistinguishability of the source). 
An electro-optic modulator-based demultiplexer (DMX) routes photons into two different spatial modes.
Subsequently, an 8-mode programmable integrated photonic circuit, realized via femtosecond laser waveguide writing on a glass substrate~\cite{Corrielli2021, Pentangelo2024} and providing a universal architecture~\cite{Clements2016}, is used to encode the various parts of the VQE. 
The ensemble VQE is implemented using a subset of six modes on the circuit: 4 computational modes ($1-4$) that form two path-encoded qubit registers, and two ancillary modes ($a_0, a_1$) required to perform the post-selected C-NOT gate operation~\cite{kok2007linear}. 
The photons are path-encoded into two separate qubits across distinct spatial modes and initialized in the ensemble of two orthogonal states $b =  \lbrace\ket{++}, \ket{--}\rbrace$, where the single qubit states $\vert{}\pm\rangle = (\ket{0} \pm \ket{1})/\sqrt{2}$ are the eigenstates of the $\hat{X}$ Pauli operator.
The variational section (outlined in blue) comprises single-qubit rotations placed both before and after a central C-NOT gate. This configuration yields four variational parameters, $\vec{\theta} = (\theta_1, \theta_2, \theta_3, \theta_4)$, which control the reflectivities of the integrated Mach-Zehnder interferometers (see the inset of Fig.~\ref{fig:8modechip}). 
The final stage of the circuit projects each single-qubit register onto the Pauli observable bases required for energy estimation. 
The energy evaluation and the subsequent parameter updates rely on single-photon counts, recorded by superconducting nanowire single-photon detectors (SNSPDs) and processed by time-to-digital converters (TDCs). 
The overall photon evolution throughout the circuit is governed by 56 reconfigurable thermo-optic phase shifters.

\subsubsection{Numerical simulation using a noisy model}
\label{sec:noisy_model}

\begin{figure}[t]
    \centering
    \includegraphics[width=1\linewidth]{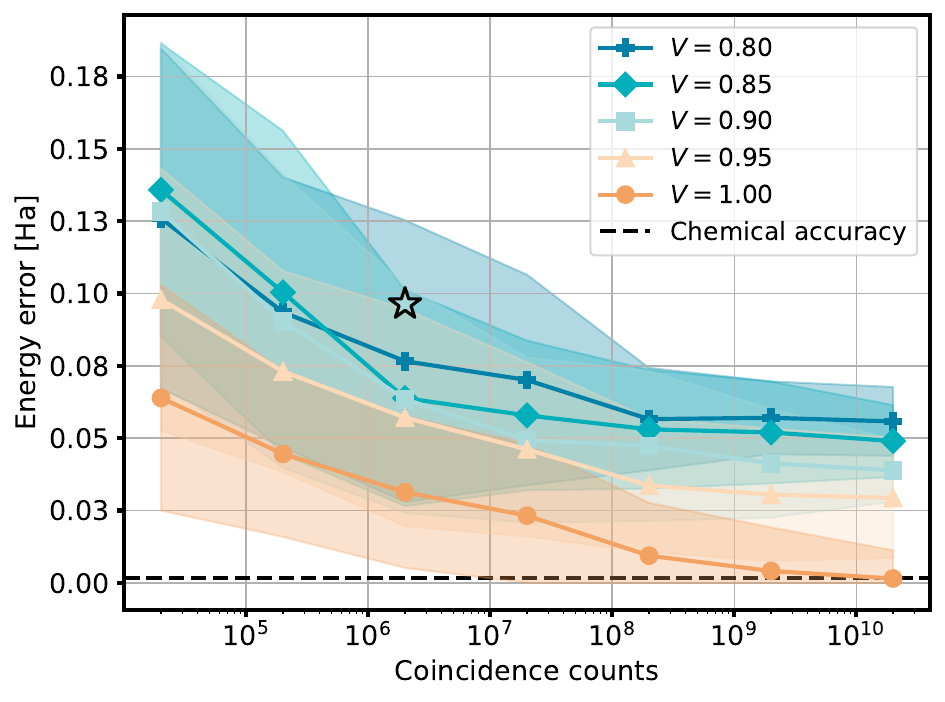}
    \caption{\textbf{Impact of noise on the convergence and accuracy of the  ensemble VQE.}
    Simulations for the energy error (distance from the theoretical minimum of the estimated cost function by the ensemble VQE) for the last iteration ($H_6$) of the Q-DFT (see Fig.~\ref{fig:SCF_LiH}), as a function of the two-photon counts for different values of HOM visibility $V$. The latter parameter quantifies the degree of photon indistinguishability, where $V=1$ corresponds to fully indistinguishable photons and $V=0$ to fully distinguishable ones. The figure displays the mean values and standard deviation bands over 100 VQE simulated runs. The star marker indicates the experimental operating conditions and the corresponding energy error achieved in the best experimental run. The chemical accuracy threshold at $0.0016$ Ha is indicated by the dashed black line.}
    \label{fig:acc_exp_vis_sim}
\end{figure}

Here we investigate the impact of experimental imperfection in the performance of the photonic ensemble VQE. The two main sources of noise are the residual single-photon partial distinguishability and the limited photon counts. 
Simulations of the experiment including a model of partial distinguishability and finite statistics showed that these two parameters actually hinder the convergence of the ensemble VQE, raising a systematic error shown in Fig.~\ref{fig:acc_exp_vis_sim}.
The simulation was conducted by initializing the photon pairs in the states of the 
$b$
ensemble for different values of photon indistinguishability (quantified by the HOM-effect visibility \cite{barzaghi2025lowloss24mode}) and number of photons detected, and averaging the final energy obtained after 100 full VQE optimization runs of the Hamiltonian $H_6$, corresponding to the final iteration of the noiseless Q-DFT loop for the optimization of the electron density of the LiH molecule (Fig.~\ref{fig:SCF_LiH}).

We recall that the sequence of the six Hamiltonians
was generated entirely offline via 
noiseless Q-DFT calculations run on a classical computer that emulates the photonic hardware.
We adopted this decoupled strategy because the unavoidable presence of noise in the current hardware conditions, when accounted for in the model, does not allow to resolve the different iterations of a fully closed, self-consistent Q-DFT loop. Indeed, as shown in Fig.~\ref{fig:SCF_LiH}, even shot noise alone can already hinder the convergence of the SCF loop.
When the photon counts are sufficiently large, the error is dominated by the partial distinguishability of photons, and a near-perfect photon indistinguishability is needed to achieve chemical accuracy (dashed black line in Fig.~\ref{fig:acc_exp_vis_sim}).
A comprehensive discussion of this limitation, including a detailed analysis of primary noise sources, such as partial photon indistinguishability, and their impact on algorithmic performance, is provided in the Methods and Appendix~\ref{app:noise}.

\subsubsection{Experimental results and chemical analysis} 

\begin{figure}
    \centering
    \includegraphics[width=1\linewidth]{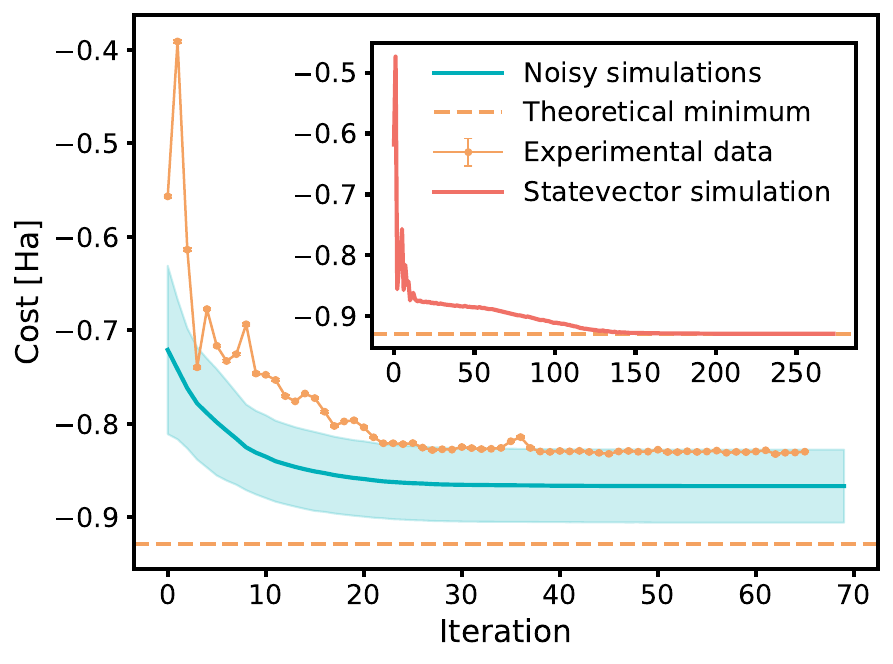}
    \caption{\textbf{Ensemble VQE results.}
    The best experimental convergence run of $H_6$ is shown. The error bars, propagated from the standard deviation of the Poissonian distribution of photon counts, are smaller than the data markers. The blue shaded band represents the standard deviation evaluated over 100 numerical simulations. The underlying theoretical model accounts for finite counting statistics (shot noise) and the effective experimental photon indistinguishability of $\sim 89\%$. The simulated shot noise matches the experimental conditions, corresponding to a detector integration time of $t=100$~s and a measured two-photon coincidence rate of $\sim 20\times10^3$ pairs per second at the output of the integrated photonic processor. Inset:  Ensemble energy [Eq.~(\ref{eq:ensE})] as a function of ensemble VQE iteration obtained via an ideal state-vector simulation. 
    }
    \label{fig:best_run_exp}
\end{figure}

For each specific Hamiltonian, the cost function is estimated by post-selecting exclusively on coincidence events in which exactly one photon is detected in the first spatial register (modes 1 and 2) and one photon in the second register (modes 3 and 4). 
This post-selection condition is strictly required to ensure the correct implementation of the protocol's logical subspace. 
Following the measurements, a classical subroutine calculates the expectation values and proposes an updated optimal set of variational parameters, $\vec{\theta}_{\text{opt}}$, by minimizing the cost function via the COBYLA algorithm. 
Representative experimental outcomes are illustrated in Fig.~\ref{fig:best_run_exp}.
As an example, we report the optimization trajectories performed on the $b$ ensemble for the final ($H_6$) Hamiltonian. 
 The results of the experimental VQE on the previous five iterations of Q-DFT are provided and discussed in Appendix~\ref{app:10s}.

Simulations with the model of the previous section that takes into account the partial photon distinguishability and finite statistics, represented by the blue line and bands in Fig.~\ref{fig:best_run_exp}, show that the results, albeit not compatible with the target, remain in agreement with theoretical predictions, given these noise sources. 
Moreover, this constitutes merely a lower bound as other sources of error, such as imperfections in the control of the phase shifters of the interferometric mesh, were not considered in the model.

\begin{figure*}
    \centering
    \includegraphics[width=0.9\linewidth]{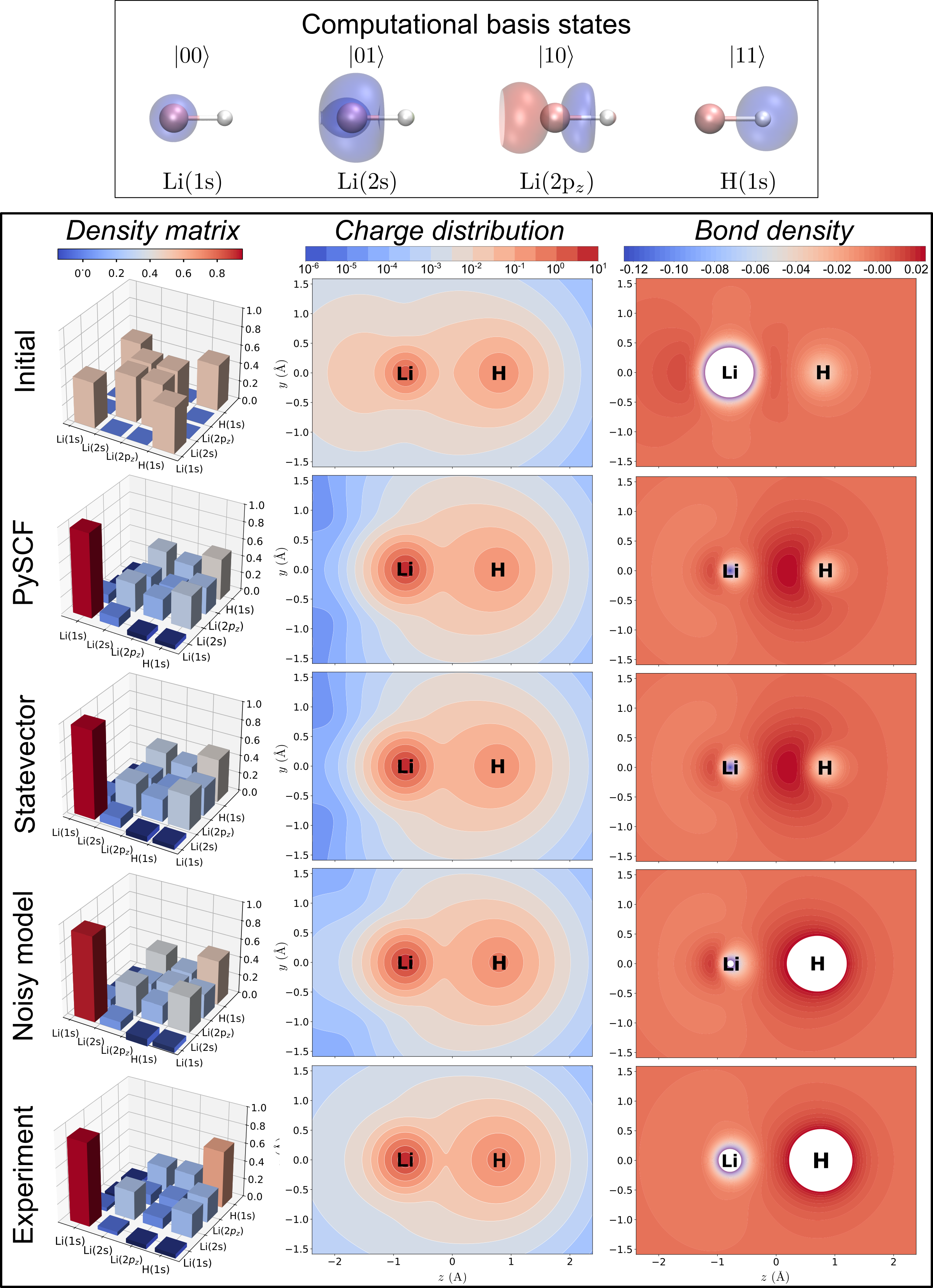}
    \caption{
    \textbf{Chemical analysis of the ensemble VQE results.}
     Orthonormal atomic orbitals corresponding to the computational basis states are shown on the first block.
     On the second block, each column shows different representations of the
     results: 
     the one-particle density matrix, the charge distribution and the bond density map, respectively.
     Each row corresponds to different subspaces spanned by the occupied molecular orbitals:
     the
     initial starting point of the ensemble VQE (before any application of the parameterized ansatz $U(\bmtheta)$), the exact PySCF reference (exact diagonalization), the noiseless statevector simulation, the noisy model simulation, and the experimental simulation, respectively.}
    \label{fig:chemistry}
\end{figure*}

Let us now turn to the chemical interpretation of the results obtained in Fig.~\ref{fig:best_run_exp}. 
In Q-DFT, the computational basis corresponds to an orthonormal set of orbitals or basis functions. In this work, we use orthonormal atomic orbitals (OAOs), obtained from L\"owdin symmetric orthogonalization of the atomic orbitals~\cite{lowdin1950non}, although other choices, such as plane waves or Daubechies wavelets~\cite{hong2022accurate} can also be used in practice. These are shown in the first block of Fig.~\ref{fig:chemistry},
and are referred to as 
Li(1s) for $\ket{00}$,
Li(2s) for $\ket{01}$,
Li(2p$_z$) for $\ket{10}$
and H(1s) for $\ket{11}$,
due to their strong resemblance
to the original (non-orthogonal) atomic orbitals. 
The linear combination of atomic orbital (LCAO) coefficients are provided in Appendix~\ref{app:orbital}.

In Fig.~\ref{fig:chemistry}, we also represent 
the one-particle density matrix, the charge and the bond density map obtained within our different simulations.
The one-particle density matrix of the KS problem
expressed in the OAO basis corresponds to the sum of the density matrices of the two optimized states, computed according to Eq.~(\ref{eq:opdm}).
Its diagonal elements encode the occupation of the OAOs by the four electrons of the system, while its off-diagonal elements describe the delocalization of the occupied molecular orbitals over the OAOs.
Turning to the charge and bond density maps, they are usually used to characterize the nature of chemical bonding in a molecule, as covalent and ionic bonds exhibit markedly different chemical and physical properties.
The first represents the molecular charge distribution that describes how the total electronic density is distributed in space. 
The second represents the bond density, or density difference, and shows the difference between the molecular charge 
distribution and the superposition of the free, spherically averaged atomic densities placed at the nuclear positions
(generally used as a guess density to start a self-consistent DFT calculation and also referred to as the independent atom model).
In the following, we investigate how the ensemble VQE optimization manages to efficiently evolve the initial states toward the correct occupied molecular orbitals and real-space density, given by the PySCF reference calculation.

Starting with the PySCF reference, the two eigenstates correspond predominantly to the non-bonding Li(1s) (core) orbital and to a bonding (valence) orbital with a large contribution from H(1s) and smaller contributions from Li(2s) and Li(2p$_z$) orbitals.
This can be seen directly from the diagonal elements of the density matrix, and from the canonical orbitals expressed in the OAO basis that are provided in Appendix~\ref{app:orbital}.
The contour map of the molecular charge distribution also
shows
a polarization that is a clear signature of ionic bonding, shaped by the combination of the H(1s), Li(2s), and Li(2p$_z$) orbitals.
This charge transfer is easier to see
on the contour map of the bond density.
Indeed, it is clear that the valence electron of lithium (blue colour at the center of lithium corresponding to a negative density difference)
moves to hydrogen, as shown by the polarized positive density difference directed toward hydrogen.

Turning to the ensemble VQE simulations, one must define a set of initial states corresponding to the number of (doubly) occupied molecular orbitals ({\it i.e.}, two for LiH).
Any of the computational basis states can be considered, as previously done in Ref.~\cite{senjean2023toward}, or any linear combinations of these states, provided 
that the resulting states remain orthonormal to each other.
In this work, we use the
ensemble $b$ defined earlier,
composed of $\ket{++} = (\ket{00}+\ket{01}+\ket{10}+\ket{11})/2$ and $\ket{--} = (\ket{00}-\ket{01}-\ket{10}+\ket{11})/2$ states
that are easy to prepare on the integrated photonic circuit.
As readily seen from the density matrix of the initial states, the electrons are completely delocalized over the diatomic molecule, with an equal contribution from all OAOs.
This is, of course, far from the optimal solution, as the OAOs of LiH differ significantly in energy.
In comparison to the charge distribution of the PySCF reference, it is clear that the initial density is not sufficiently localized on the lithium atom.
This missing occupation can also be observed on the bond density map. 
Indeed, there is a large difference of density at the center of the lithium atom, while the difference is much smaller elsewhere in comparison.
Being different to the independent atom model is not an issue, as the ensemble VQE ansatz should be expressive enough to converge to the targeted subspace spanned by the PySCF reference states.
However, in practice it could hinder convergence, as the optimization landscape is known to be prone to many local minima or barren plateaus, especially when using hardware-efficient ansatz~\cite{larocca2025barren}.

Let us now consider the results obtained after optimizing the occupied molecular orbital subspace within ensemble VQE.
It is worth noting that the noiseless statevector simulation reproduces almost exactly the PySCF reference results.
Hence, the single-layer $R_Y$C-NOT ansatz is expressive enough,
and the optimization does not end up in a local minimum.
However, when turning to the best run of the noisy-model simulations (the one giving the lowest ensemble energy), 
visible differences emerge. 
Interestingly, 
the molecular charge distribution retains characteristics similar to those of the 
statevector simulation, but with a much higher density difference centered on the atoms (negative difference on the lithium and positive difference on the hydrogen), as can be seen by the white circles that are beyond the values depicted by the color bar.
In addition, the density between the atoms is lower, displaying a smaller bonding contribution compared to the statevector simulation.
The polarization due to the Li(2p$_z$) orbital can still be recognized on the map, especially from the positive difference of density on the left of the lithium atom.

As expected, the experimental result deviates even further from the PySCF reference results,
although it is still much better than the initial states that are really far from the optimal solutions.
For the averaged experimental density matrix $\gamma_{\mathrm{exp}}$,
whose fidelity with the averaged theoretical (statevector) density matrix $\gamma_{\mathrm{th}}$ is $\mathcal{F} = \mathrm{Tr}^2 \left(\sqrt{\sqrt{\gamma_{\mathrm{th}}} \gamma_{\mathrm{exp}} \sqrt{\gamma_{\mathrm{th}}}}\right) = 90.5\%$
and Frobenius norm is 0.30, the H(1s) orbital exhibits a larger occupation than the reference, while the orbitals Li(1s) and Li(2p$_z$) have smaller occupations.
The lack of contribution from Li(2p$_z$) can be observed in the contour map, which depicts much more spherical isosurfaces compared to the polarized reference or statevector ones, or even from the noisy model simulation.
Hence, although the charge transfer is still reproduced by the experiment, it dominates over the bonding character, meaning that the states are contaminated by virtual states that exhibit non-bonding or anti-bonding character.
This can be seen by the white circles being even larger than for the noisy model simulation and the absence of charge density in between the atoms.
To recover the proper ionic bonding, error mitigation should be considered, as well as future hardware improvements on photon indistinguishability and optical losses.

Overall, although the sources of error beyond those included in the noisy model (see Sec.~\ref{sec:noisy_model}) are too significant to provide quantitatively accurate results, the noisy-model simulations, incorporating partial photon distinguishability and photon-counting statistics, show promise for NISQ applications of Q-DFT to quantum chemistry.

\section{Conclusion}\label{sec13}
In this work, we presented an experimental demonstration of Q-DFT mapped onto a programmable integrated photonic circuit. Specifically, we implemented the ensemble extension of the VQE on Kohn--Sham Hamiltonians for the lithium hydride molecule. Our experimental results demonstrated the convergence of the VQE optimization routine on the physical device. Minimizing the energy within the subspace spanned by the ground and first excited states allowed us to evaluate the corresponding state properties and explicitly compute the molecular electron density.
Although the statevector simulations yielded results that were almost indistinguishable from the reference results, the experimental results were not yet 
sufficient to accurately describe the ionic bond in the LiH diatomic molecule.
Nevertheless, the noisy-model simulation successfully obtained qualitatively meaningful results, demonstrating that even modest improvements to the experimental setup could enable relevant electronic-structure simulations on photonic devices.

These findings indicate a clear trajectory for future hardware improvements, pinpointing photon indistinguishability and optical losses as the primary hardware requirements to address. Rapid advancements in single-photon sources technologies are continuously enhancing their brightness and photon indistinguishability \cite{Senellart2017,Ding2025}. Concurrently, the development of integrated photonic circuits is steadily reducing propagation losses while reaching greater circuit scale, depth and complexity \cite{Alexander2025, barzaghi2025lowloss24mode, dimicco2026}. 
Beyond these near-term improvements, scaling this photonic protocol to larger molecular systems requires overcoming the constraints of probabilistic multi-qubit operations, such as the current C-NOT implementation. It is well established that optical nonlinearities at the single-photon level are strictly necessary to achieve deterministic logic gates and to generate the highly entangled states required for VQE encodings \cite{kok2007linear}. While engineering direct single-photon nonlinearities remains experimentally challenging, current state-of-the-art methods address this requirement through various pathways. One  approach relies on measurement-induced nonlinearities \cite{Scheel2003, Sparrow2018} combined with adaptive operations~\cite{Hoch2025, Rodari2026}. Furthermore, solid-state emitters such as quantum dots can be exploited either to deterministically generate multi-photon entangled states \cite{Huet2025, huet2026industryready} or to mediate nonlinear photon-photon interactions, as recently demonstrated in a proof-of-concept nonlinear quantum photonic simulator of molecular vibrational dynamics~\cite{Nielsen2025}.

Such nonlinear and highly reconfigurable integrated photonic platforms will unlock the full potential of qubit-efficient photonic algorithms for quantum chemistry.
This paves the way for 
extending the reach of photonic Q-DFT from small benchmark molecules such as LiH to systems of technological relevance, such as lithium-containing compounds for next-generation batteries.
A particularly relevant target is the modeling of extended electrochemical interfaces as a function of the applied potential, which would allow for a better understanding of the complex chemistry of the solid electrolyte interphase (SEI), at the core of the present limitations in fast charging, cyclability and ageing of Li batteries~\cite{SEI25,SE26}.

\section{Methods}\label{sec:methods}


\subsection{Theoretical methodology}

\subsubsection{Quantum density functional theory and qubit-efficient mapping}

Q-DFT is the quantum algorithm analogue of KS-DFT, the workhorse method in quantum chemistry and condensed matter physics on classical computers.
The undeniable success of KS-DFT relies on the mapping of the complex many-body quantum electronic problem into an effective and density-dependent one-body problem~\cite{hktheo,KS},
\begin{eqnarray}
\left( \hat{T} + \hat{v}^{\rm KS}[n] \right) \ket{\Phi^{\rm KS}} = \mathcal{E}^{\rm KS} \ket{\Phi^{\rm KS}},    
\label{eq:KS}
\end{eqnarray}
from which the exact ground-state energy and ground-state electronic density can be recovered in principle.
In Eq.~(\ref{eq:KS}), \(\hat{T}\) is the kinetic energy operator, \(\ket{\Phi^{\rm KS}}\) is a single Slater determinant, and \(\hat{v}^{\rm KS}[n] = \hat{v}^{\rm ext} + \hat{v}^{\rm H}[n] + \hat{v}^{\rm xc}[n]\) is the density-dependent KS potential operator containing the external potential (i.e., the ion-electron interaction), the trivial Hartree potential and the exchange and correlation (XC) potential, respectively.
The latter is a one-body quantity that effectively accounts for all non-trivial contributions arising from the two-body electron-electron interaction.
As it is a one-body problem, it can be rewritten as
\begin{eqnarray}
    \hat{h}^{\rm KS}[n] \ket{\varphi_k} = \varepsilon_k \ket{\varphi_k},
\end{eqnarray}
where $\lbrace\ket{\varphi_k}\rbrace$
are the KS orbitals in $\ket{\Phi^{\rm KS}}$.
In second quantization, when described
in a finite computational basis-set of
$N_{\rm bf}$ basis functions $\lbrace \ket{\psi_\mu}\rbrace$, we have, in the spin-restricted formalism,
  \begin{eqnarray}\label{eq:H_mono}
    \hat{h}^{\rm KS}[n] &= &\sum_{\mu=1,\nu=1}^{N_{\rm bf}}  \sum_{\sigma=\lbrace \uparrow, \downarrow\rbrace} h_{\mu\nu}^{\rm KS}[n]  \left( \hat{c}^\dagger_{\mu\sigma} \hat{c}_{\nu\sigma} + {\rm h.c.} \right),
  \end{eqnarray}
where $\hat{c}^\dagger_{\mu\sigma}$ ($\hat{c}_{\mu\sigma}$) refers to the creation (annihilation) operator of an electron of spin $\sigma$ in orbital $\ket{\psi_{\mu}}$, respectively,
and
$h_{\mu \nu}^{\rm KS}$ contains both the kinetic contributions and the KS potential. 
The electronic density,
where \(N_{\rm occ}\) denotes the number of doubly-occupied orbitals,
reads
\begin{eqnarray}
    n(\bfr) &=& 
    2\sum_{k=1}^{N_{\rm occ}}
    | \varphi_k(\bfr) |^2 \nonumber \\
    &=& 
    2\sum_{k=1}^{N_{\rm occ}}
    \psh{\bfr}{\varphi_k}\psh{\varphi_k}{\bfr}, \nonumber \\
&=&
    2\sum_{k=1}^{N_{\rm occ}}
    \sum_{\mu,\nu}^{N_{\rm bf}}
    \psh{\bfr}{\psi_\mu}\psh{\psi_\mu}{\varphi_k}\psh{\varphi_k}{\psi_\nu}\psh{\psi_\nu}{\bfr},\label{eq:density}
\end{eqnarray}
and is used to compute a new KS potential, so that the KS problem
in Eq.~(\ref{eq:KS}) is solved self-consistently.
After convergence,
the ground-state energy of the physical system can be recovered
as follows,
\begin{eqnarray}
    E_0 = \mathcal{E}^{\rm KS} + E_{\rm Hxc}[n^{\Phi^{\rm KS}}] - \int v^{\rm Hxc}[n^{\Phi^{\rm KS}}](\bfr) \, n^{\Phi^{\rm KS}}(\bfr) \, {\rm d}\bfr,
\end{eqnarray}
where \(\mathcal{E}^{\rm KS} = 2\sum_{k=1}^{N_{\rm occ}} \varepsilon_k\).
While the theory is in-principle exact,
the XC functional remains unknown despite plenty of different flavors of approximations~\cite{marques2012libxc}.

Solving the KS problem on a quantum computer consists in encoding
the $N_{\rm bf}$ orthonormal orbitals $\lbrace \ket{\psi_\mu}\rbrace$
on the states of a quantum register.
We use the qubit-efficient (also called binary) mapping~\cite{shee2022qubit,senjean2023toward,knapik2026quantum},
so that the $N_{\rm bf}$ basis functions are
mapped to $M = \log_2(N_{\rm bf}) = 2$ qubits
only and the KS orbitals are expressed as a (normalized) linear combination of the basis functions
\begin{eqnarray}
    \ket{\varphi_k} &=& \sum_\mu^{N_{\rm bf}} c_{\mu k} \ket{\psi_\mu}\nonumber \\
    &=& \sum_\mu^{2^M} c_{\mu k} \ket{\mu},
\end{eqnarray}
where $\ket{\mu}$ is the binary string corresponding to the integer $\mu$.
Then,
the operator $\hat{h}^{\rm KS}[n]$
is expressed as a linear combination of unitaries, such as Pauli strings,
and can be solved by any quantum algorithm
able to extract the first $N_{\rm occ}$
excited states
required to compute the electronic density~\cite{senjean2023toward}.
The latter is obtained by
estimating the density matrix elements
\begin{eqnarray}
    \gamma_{\mu\nu}
    =  \sum_{k=1}^{N_{\rm occ}}\psh{\psi_\mu}{\varphi_k}\psh{\varphi_k}{\psi_\nu} 
    \label{eq:opdm}
\end{eqnarray}
on the quantum device
state tomography
and the projection
onto the real-space grid is performed on the classical computer, as shown in Eq.~(\ref{eq:density}).
In this work,
we used orthogonal atomic
orbitals, obtained
by L\"owdin symmetric orthogonalization~\cite{lowdin1970nonorthogonality} of the atomic orbitals of the minimal STO-3G basis,
keeping only the 1$s$, 2$s$ and 2$p_z$
orbitals of lithium, and the 1$s$
of hydrogen ($2p_x$ and $2p_y$ do not participate due to symmetry reasons).
Hence, only 2 qubits are needed
to compute the energy of the LiH molecule in the minimal basis at the KS-DFT level of theory.
Since LiH has four electrons, they occupy two KS orbitals, thus requiring a quantum solver for both the ground and the first-excited state.
In this work, we considered the ensemble variational quantum eigensolver.

\subsubsection{Ensemble variational quantum eigensolver}
In order to compute both the ground and the first-excited state energies,
we use the ensemble VQE based on the generalization of the Rayleigh-Ritz principle~\cite{gross1988rayleigh,nakanishi2019subspace,yalouz2021state,rajamani2025equi,soule2026unified}.
For two states $A$ and $B$ with 
weights $w_A$ and $w_B$,
the ensemble energy to be minimized
reads
\begin{eqnarray}
    E^\bfw(\bmtheta) &=& \sum_{I=\lbrace A,B \rbrace}w_I \,\bra{\varphi_I(\bmtheta)}\hat{h}^{\rm KS}\ket{\varphi_I(\bmtheta)}
    \nonumber \\
    &=&
    \sum_{I=\lbrace A,B \rbrace}w_I \,\varepsilon_I(\bmtheta).
     \label{eq:ensE}
\end{eqnarray}
This energy can be related to the
energy of the auxiliary KS system in Eq.~(\ref{eq:KS}), {\it i.e.,} $\mathcal{E}^{\rm KS} = 2(\varepsilon_0 + \varepsilon_1)$,
if $w_A = w_B = 2$ and $\ket{\varphi_A}$ and $\ket{\varphi_B}$ span the same subspace as
the first two eigenstates of $\hat{h}^{\rm KS}$, as $\mathcal{E}^{\rm KS}$ is invariant by rotation of $\ket{\varphi_0}$ and $\ket{\varphi_1}$
These conditions are
exactly fulfilled by the `equi-weights' formulation of ensemble VQE,
which is therefore enough to get the energy and the electronic density in Q-DFT, thus avoiding optimization issues when considering ordered weights~\cite{rajamani2025equi}.
The states in Eq.~(\ref{eq:ensE})
are prepared on parametrized quantum circuits as follows,
\begin{eqnarray}
       \ket{\varphi_A(\bmtheta)} &=& U(\bmtheta) \ket{\chi_A} ,\nonumber\\
    \ket{\varphi_B(\bmtheta)} &=& U(\bmtheta) \ket{\chi_B} , 
\end{eqnarray}
where $\ket{\chi_A}$ and $\ket{\chi_B}$
are orthonormal easy-to-prepare initial states,
taken as $\ket{\chi_A} = \ket{++}$ and $\ket{\chi_B} = \ket{--}$, corresponding to superpositions of all the computational basis states (orthonormal atomic orbitals $\lbrace \ket{\psi_\mu} \rbrace$). 
Any other choice can be made as long as they remain orthonormal, as the same unitary operation is applied on both states to preserve the orthonormality).
The $R_Y$C-NOT hardware efficient ansatz~\cite{kandala2017hardware},
\begin{eqnarray}
U(\bmtheta) &=& \prod_{m=1}^M R_{Y,m}(\theta_{m_y}^0)\prod_{n=1}^{\Gamma} U_n^{\rm ENT}(\bmtheta^n),
\end{eqnarray}
were used,
with a single layer $\Gamma = 1$ and a number of qubits $M=2$.
The entanglement unitary block reads
\begin{eqnarray}
U_n^{\rm ENT}(\bmtheta^n) = 
\prod_{m=1}^{M-1} {\text{C-NOT}}_{m(m+1)} \prod_{m=1}^M R_{Y,m}(\theta_{m}^n).
\end{eqnarray}
The gradient-free COBYLA 
optimizer were considered in this work.

\subsection{Photonic-based quantum circuits}
\subsubsection{Dual-rail encoding}

In the dual-rail encoding~\cite{kok2007linear,Wang2019},
a logical qubit is represented by a single photon
occupying one of two optical modes,
conventionally labelled $a$ and $b$.
Denoting by $\ket{n_a, n_b}$
the Fock state with $n_a$ ($n_b$) photons in mode $a$ ($b$),
the computational-basis states of a single qubit read
\begin{eqnarray}
    \ket{0}_L &\equiv& \ket{1,0} = \hat{a}^\dagger\ket{\rm vac}, \nonumber \\
    \ket{1}_L &\equiv& \ket{0,1} = \hat{b}^\dagger\ket{\rm vac},
    \label{eq:dualrail}
\end{eqnarray}
where $\hat{a}^\dagger$ and $\hat{b}^\dagger$ are the bosonic creation operators
of the two modes and $\ket{\rm vac}$ is the vacuum.
A key advantage of this encoding is that photon loss
maps a valid codeword to the vacuum state $\ket{0,0}$,
which lies outside the computational subspace and can therefore be detected
as an erasure error.

Single-qubit rotations are realized with passive linear optical elements.
A phase shifter on mode $a$ imparts a relative phase $\phi$
and acts on the creation operators as~\cite{kumar2021}
\begin{eqnarray}
    PS(\phi) = \begin{pmatrix} e^{i\phi} & 0 \\ 0 & 1 \end{pmatrix},
    \label{eq:phaseshifter}
\end{eqnarray}
while a reconfigurable beam splitter with transmissivity parameter $\theta$, as implemented in integrated photonics with two balanced directional couplers separated by an internal phase shifter, applies the transformation
\begin{eqnarray}
    BS(\theta) = i e^{i \frac{\theta}{2}} \begin{pmatrix} \sin\frac{\theta}{2} & \cos\frac{\theta}{2} \\ \cos\frac{\theta}{2} & -\sin\frac{\theta}{2} \end{pmatrix}.
    \label{eq:beamsplitter}
\end{eqnarray}
The fundamental building block of our programmable circuit is the MZI, consisting of an external phase shifter $PS(\phi)$ on one input mode, followed by a reconfigurable beam splitter $BS(\theta)$. This configuration implements the parametrized unitary
\begin{equation}
 \text{MZI}(\phi,\theta) = BS(\theta)PS(\phi) 
\label{eq:MZI}
\end{equation}
which is the optical unit of universal multi-mode interferometers~\cite{Clements2016}, and, together with an output phase, can realize an arbitrary $\text{SU}(2)$ transformation on a single dual-rail qubit.

\subsubsection{Experimental set-up} 
A single-photon source based on a semiconductor quantum dot (QUANDELA source) in the resonance fluorescence configuration is excited by a pulsed laser at the repetition rate of $\sim 79$ MHz. The single-photon purity is evaluated via a Hanbury Brown and Twiss experiment, measuring a second-order correlation function of $g_2(0) \sim 3\%$. An electro-optic modulator-based demultiplexer (QUANDELA DMX) routes photons into to 2 different channels to perform a two-photon experiment. 
The photons indistinguishability is estimated by measuring the visibility of a Hong-Ou-Mandel experiment \cite{HOM, Bouchard_2021} on the photonic circuit, yielding $V^\mathrm{HOM} \sim 89\%$. 
Single-photon counts are recorded by SNSPDs (Opus One system) and processed by a TDC device (PicoQuant MultiHarp). 

The integrated device is fabricated in glass via the femtosecond laser writing technique. Its internal structure follows the universal architecture of Ref.~\cite{Clements2016}, in which the MZIs defined in Eq.~\eqref{eq:MZI} are arranged in a rectangular mesh (see also Fig.~\ref{fig:8modechip}). A total of 56 heaters provide full control over the phases $\phi$ and $\theta$ of each MZI.

\subsection{Open-source software simulations.} Numerical simulations have been performed to motivate and guide the experimental investigation. We used \texttt{qlass} \cite{Labib2026}, an open-source software developed to enable the translation of a quantum chemical program into a program compatible with the experiment (photonic circuit characteristics and setup). \texttt{qlass} is used to run ensemble VQE simulations for different molecular Hamiltonians. \texttt{Perceval} is used to define the optical circuit and noise model \cite{heurtel2023perceval} as well as a custom noise model. 







\section*{Acknowledgements}




We acknowledge support from the European Union through the QLASS project (EU Horizon Europe grant agreement 101135876), by ICSC – Centro Nazionale di Ricerca in High Performance Computing, Big Data and Quantum Computing, funded by European Union – NextGenerationEU and from PNRIC ‘Q-SUD’ (1\_16868501004\_0001741), Action 1.1.3b ‘PIMIQ’ (1741\_QIIR113b\_00050).  
Fabrication of the femtosecond laser-written photonic circuit was partially performed at PoliFAB, the micro and nano-fabrication facility of Politecnico di Milano (https://www.polifab.polimi.it). F.C. and R.O. wish to thank the PoliFAB staff for the valuable technical support.
B.S. also thanks the French National Research Agency (ANR) under the project ANR-23-CE29-0004-01 and ANR-23-PETQ-0006. 

\textbf{Conflict of interest/Competing interests:} The authors declare no competing interests. 

\textbf{Data and code availability:}
The code supporting the results and the data of the different simulations are available upon request.

\newcommand{\Aa}[0]{Aa}

\onecolumngrid

\appendix

\section{Linear combination of Pauli strings for each Q-DFT iteration}\label{app:pauli}

For the LiH molecule within the minimal basis set (keeping only the 1s, 2s, and 2p$_z$ orbitals of lithium
and the 1s orbital of hydrogen),
only $\log_2(4) = 2$
qubits are needed to compute its ground-state energy
at the KS-DFT level.
In the main text, we showed that only 6 iterations are required to converge the self-consistent Kohn--Sham loop within the noiseless Q-DFT simulation.
The six (density-dependent) Kohn--Sham Hamiltonians are given here as linear combinations of Pauli strings:

\begin{table*}[!ht]
    \centering
    \begin{tabular}{l cccccc}
        \toprule
        Pauli Term & $H_1$ & $H_2$ & $H_3$ & $H_4$ & $H_5$ & $H_6$ \\
        \midrule
        II & $-0.44216633$ & $-0.36108472$ & $-0.38614239$ & $-0.38221333$ & $-0.38213293$ & $-0.38214930$ \\
        IX & $-0.24589430$ & $-0.24059126$ & $-0.24216967$ & $-0.24205310$ & $-0.24207902$ & $-0.24208486$ \\
        IZ & $-0.36997974$ & $-0.43150506$ & $-0.40039876$ & $-0.40161551$ & $-0.40149067$ & $-0.40144943$ \\
        XI & $-0.07340748$ & $-0.07147771$ & $-0.07071070$ & $-0.06994070$ & $-0.06991539$ & $-0.06991077$ \\
        XX & $+0.00097788$ & $+0.01069059$ & $+0.00680530$ & $+0.00603800$ & $+0.00599192$ & $+0.00598300$ \\
        XZ & $+0.07417185$ & $+0.07981971$ & $+0.07767017$ & $+0.07767811$ & $+0.07766414$ & $+0.07765853$ \\
        YY & $+0.03523796$ & $+0.04986396$ & $+0.04396768$ & $+0.04347730$ & $+0.04343264$ & $+0.04342284$ \\
        ZI & $-0.44476681$ & $-0.50711419$ & $-0.47647139$ & $-0.47938523$ & $-0.47941926$ & $-0.47941549$ \\
        ZX & $-0.04444575$ & $-0.05091020$ & $-0.04831784$ & $-0.04853032$ & $-0.04850277$ & $-0.04849506$ \\
        ZZ & $-0.41780773$ & $-0.38530931$ & $-0.40068126$ & $-0.40110755$ & $-0.40114839$ & $-0.40116524$ \\
        \bottomrule
    \end{tabular}
    \caption{Coefficients of Pauli terms for Hamiltonians $H_1$ through $H_6$.}
    \label{tab:Hamiltonians}
\end{table*}

In the main text, only the last iteration, corresponding to $H_6$, was studied experimentally
with an exposure time of 100 seconds.

\section{Results with $t = 10s$ exposure}\label{app:10s}

\begin{figure*}
    \centering
    \includegraphics[width=\linewidth]{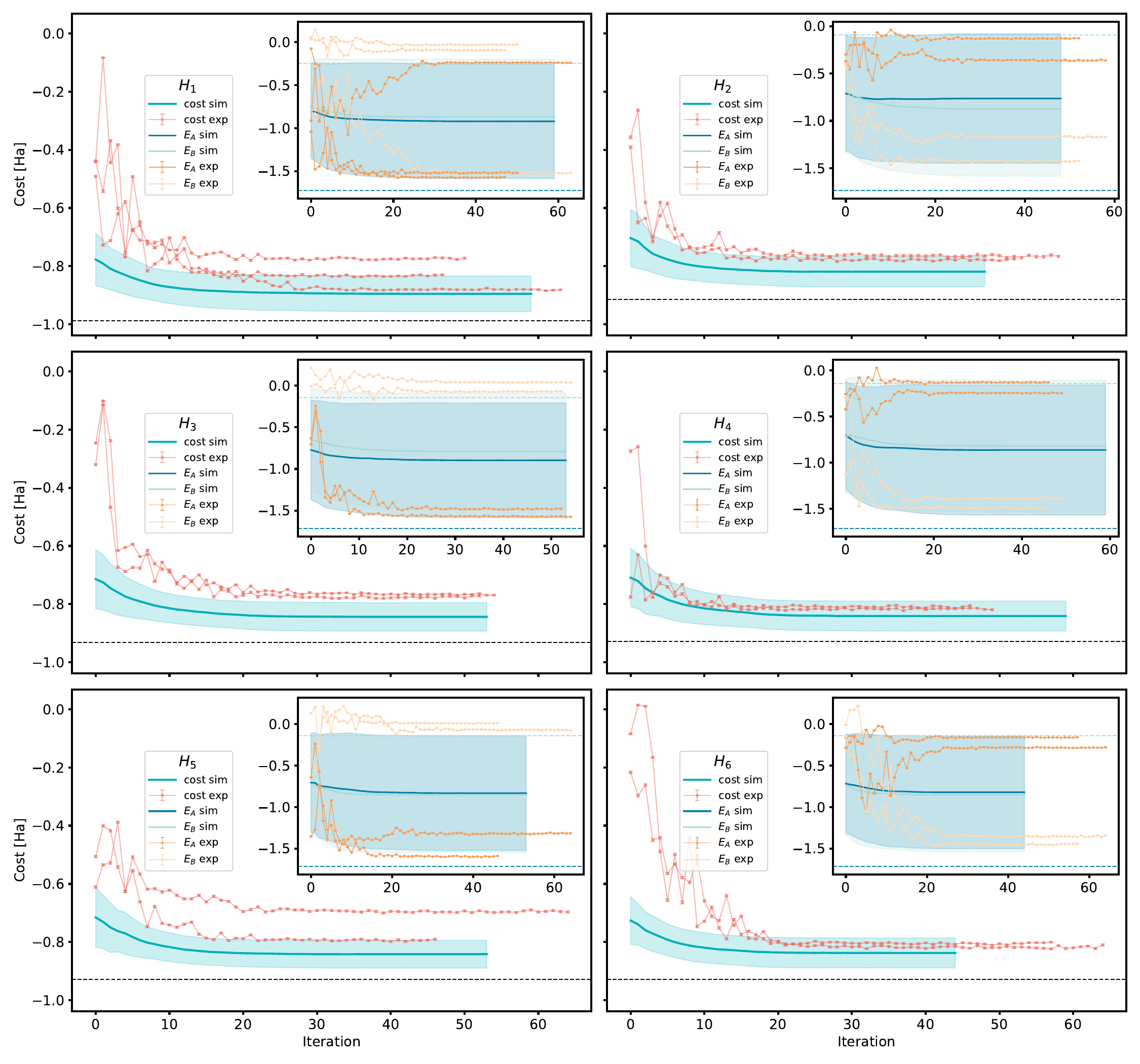}
    \caption{Experimental runs with exposure $t=10$ s. 
    The horizontal dashed dark line corresponds to the optimal value of the cost function ({\it i.e.} the ensemble energy).
    The insets show the individual energies of the two orthogonal ansatz states during the optimization.
    The horizontal dashed blue lines correspond to the ground- and first-excited state energies obtained by exact diagonalization.
    The error bars, propagated from the standard deviation of the poissonian distribution of photon counts, are too small to be visible.
    The blue bands represent the standard deviation interval for 100 simulations with a model including partial distinguishability of the photons and shot noise.}
    \label{fig:exp_10s}
\end{figure*}

Fig.~\ref{fig:exp_10s} shows the convergence of the ensemble VQE algorithm on the 6 Hamiltonians given in Table~\ref{tab:Hamiltonians}, with an exposure of 10 s.
As already discussed in the main text, a systematic error
arises due to partial distinguishability and exposure time.
The experimental results are usually slightly above the noisy model simulations, due to other sources of errors that are not included in the model, such as imperfections in the chip, given by errors in the implementation of the phases in the phaseshifters or in the directional couplers which are not perfectly balanced.
In addition to the ensemble energy, the individual energies of the two orthogonal ansatz states are shown in the insets of Fig.~\ref{fig:exp_10s}.
Let us stress that these energies do not have to be close to the eigenstate energies shown by the horizontal dashed blue lines, as only the ensemble energy is meaningful in the equi-weight ensemble VQE algorithm used here.
Indeed, the ensemble energy is invariant by rotation of the two orthogonal ansatz states,
such that only the subspace formed by the two is relevant.
This is why the noisy model simulations (in blue) show no big difference between the two orthogonal ansatz state energies,
that remain between the ground- and first-excited state energies.
Interestingly, the experimental results behave quite differently, such that the energies of the two circuits seem to converge toward the eigenstate energies.
One possible explanation for this effect is that the parameter landscape is impacted by the experimental noise in such a way that the eigenstates are easier to reach than any other states.
A deeper investigation of this feature, which can be of interest to design better optimization scheme within VQE in the NISQ era, is left for future work.

\section{Simulations with noise}\label{app:noise}

\begin{figure*}
    \centering
    \includegraphics[width=1\linewidth]{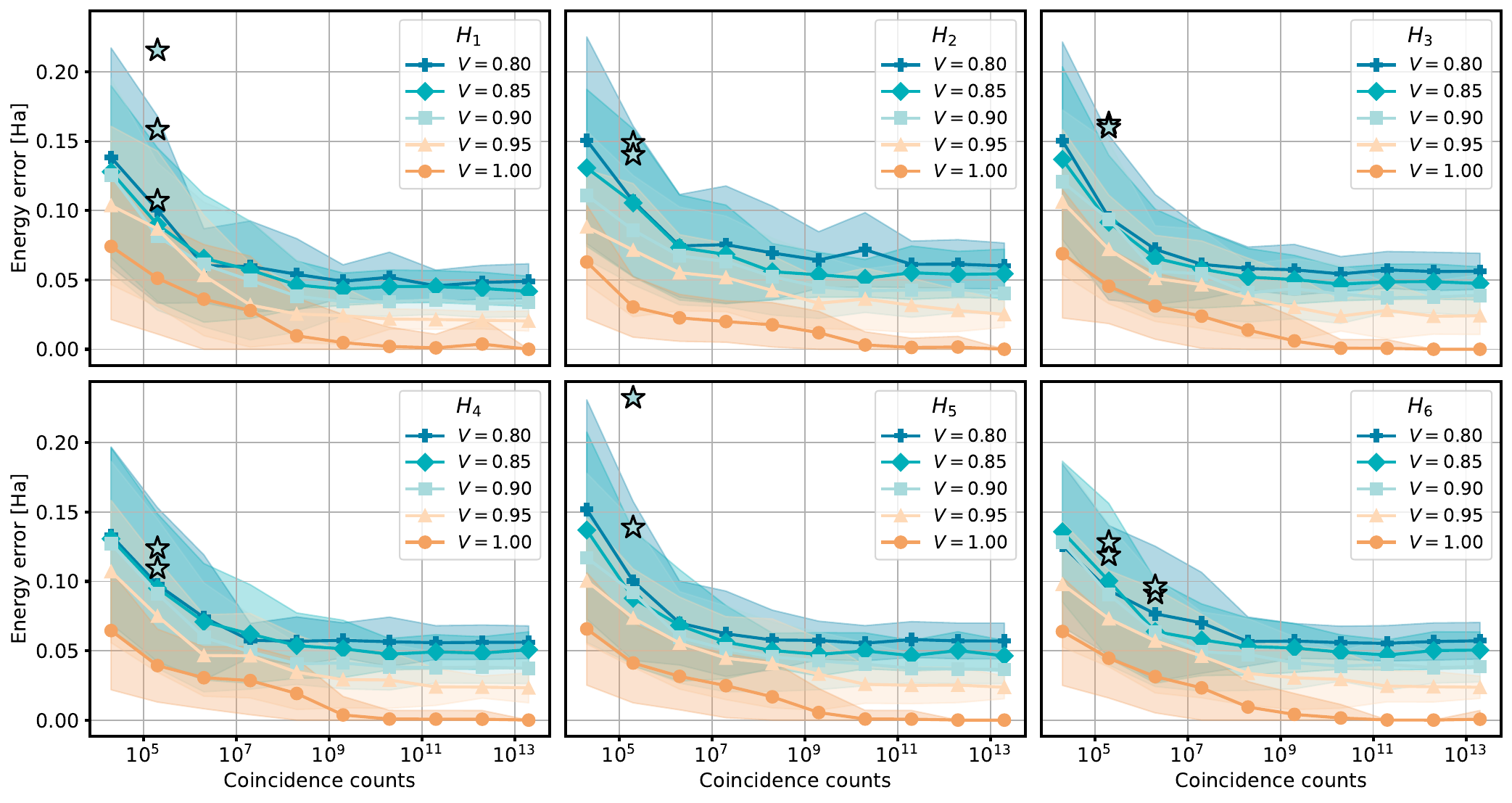}
    \caption{\textbf{Simulation with finite statistics and partial distinguishability.} The figure shows the final energy error (distance from the expected value) as a function of coincidence counts. Different colors represent different visibilities. The lines show the mean over 100 runs, while the bands show the standard deviation. Star markers represent the experimental runs at 10 s and, for $H_6$ only, 100 s exposure. 
    }
    \label{fig:noisy_sim_all_H}
\end{figure*}

In order to account for the gap between the experimental runs and the statevector simulations, simulations can be done taking into consideration two main sources of error: the partial distinguishability of photons, quantified by the Hong-Ou-Mandel visbility $V$, and the finite statistics, i.e. the low photon counts measured in the laboratory.
Visibility $V=1$ corresponds to fully indistinguishable photons, whereas $V=0$ corresponds to completely distinguishable photons which do not interact. In the experiment, the visibility achieved between the two photons injected into the integrated circuit was close to $V\sim0.89$.
The low photon counts instead depend on the measured coincidence rate at the detectors, after the intrinsic losses in the setup, and the exposure time. Fig. \ref{fig:noisy_sim_all_H} shows the final energy error (distance from the expected value) as a function of coincidence counts, for different values of visibility, obtained from an average over 100 runs with standard deviation bands. The distribution of the experimental runs, represented by star markers, is broadly in agreement with the simulation, with an additional offset due to the aforementioned sources of error not accounted for by this simple model. 
It can be observed that below a certain threshold, roughly about $10^9-10^{10}$ coincidence counts, the error is dominated by the low statistics, while improving the visibility does not necessarily yield a significative advantage in the accuracy. On the other hand, above the threshold, the final accuracy enters a plateau region where the error is dominated by the visibility of the photons, and a very high indistinguishability is required to achieve chemical accuracy, together with extremely high photon counts.

\section{Molecular orbital analysis}\label{app:orbital}

In Table~\ref{tab:LCAO}, we provide the expansion of the orthogonal atomic orbitals (OAOs)
in terms of the original non-orthogonal atomic orbitals, 
also referred to as the linear combination of atomic orbital (LCAO) coefficients.
Atomic orbitals are denoted by
1s$_{\text{Li}}$, 2s$_{\text{Li}}$,
and 2p$_{z\text{Li}}$ for lithium and 1s$_\text{H}$ for hydrogen.
As expected
from the L\"owdin symmetric orthogonalization of the atomic orbitals, 
each of
the resulting OAOs
corresponds mainly to one given atomic orbital,
which is why we named
them Li(1s), Li(2s), Li(2p$_z$) and H(1s).
Note that this transformation matrix has to be used to compute the real-space density (and thus the Hartree-exchange-correlation potential), as only the atomic orbitals are expressed in real-space.
It was also used to plot the charge and bond density maps in Fig.~6 of the manuscript.
Of course, this transformation could be bypassed by directly using an orthonormal basis expressed in real-space, such as plane waves.
\begin{table*}
    \centering
    \begin{tabular}{l cccc}
        \toprule
        basis & 1s$_{\text{Li}}$ & 2s$_{\text{Li}}$ & 2p$_{z_{\text{Li}}}$ & 1s$_{\text{H}}$ \\
        \midrule
        $\ket{00} \equiv$ Li(1s) & 1.023 & -0.129  & -0.005 & 0.006 \\
        $\ket{01}\equiv$ Li(2s) &-0.129 & 1.112  & 0.114 &-0.28 \\
        $\ket{10}\equiv$ Li(2p$_z$) & -0.005 & 0.114  & 1.146 &-0.36 \\
        $\ket{11}\equiv$ H(1s) & 0.006 & -0.28  & -0.36 & 1.234 \\
    \end{tabular}
    \caption{LCAO coefficients for the OAO orbitals, corresponding to the computational basis for our integrated photonic circuits.}
    \label{tab:LCAO}
\end{table*}

\begin{figure*}
    \centering
    \includegraphics[width=1\linewidth]{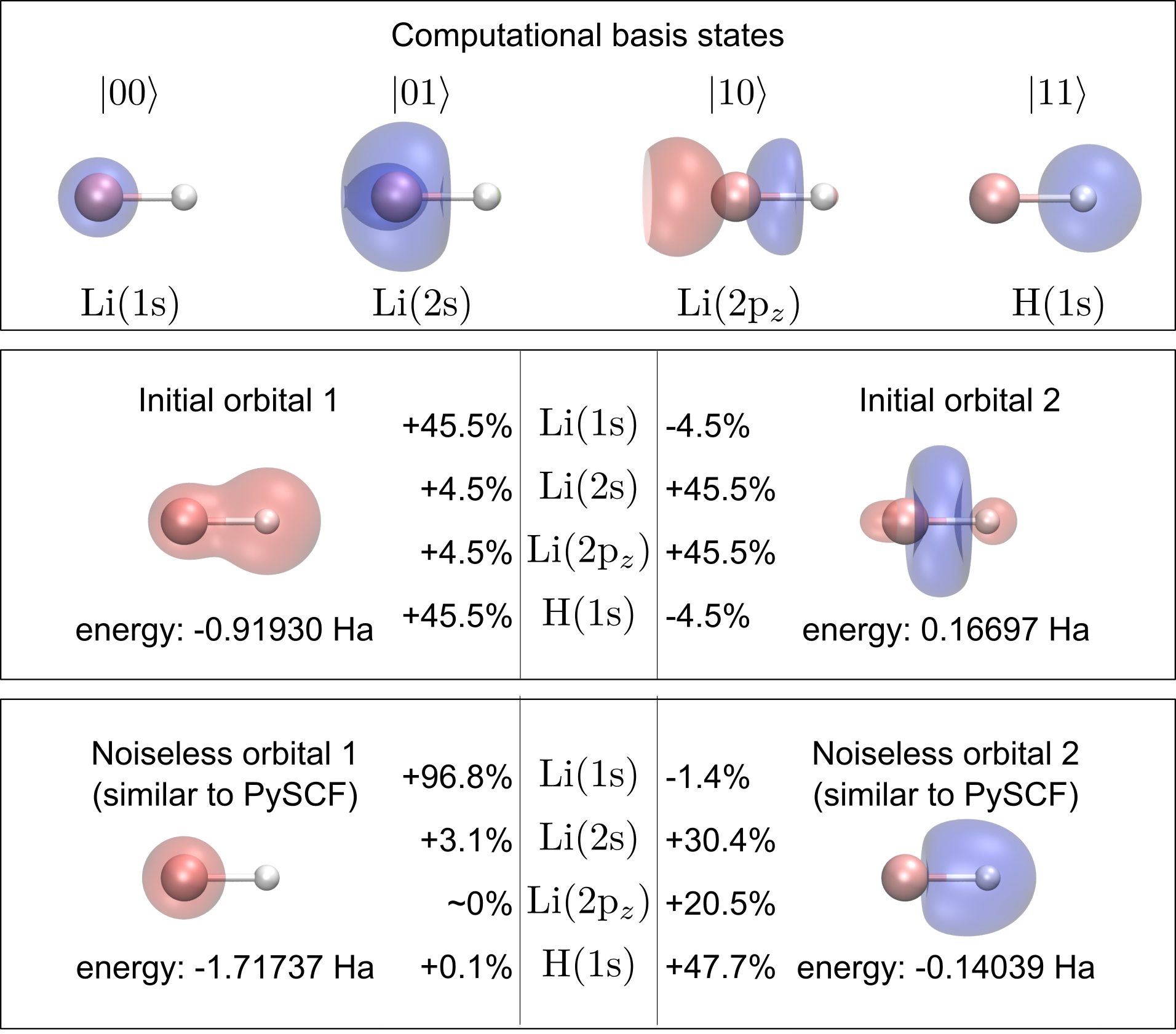}
    \caption{\textbf{Contribution of the OAOs to the canonical molecular orbitals.}
    Canonical orbitals obtained by diagonalizing the 2x2 Fock matrix in the subspace spanned by the occupied orbitals, for the initial orbitals $\ket{++}$ and $\ket{--}$ (middle panel),
    and from the converged states of the noiseless simulation $\ket{\Psi_A(\bmtheta)}$ and $\ket{\Psi_B(\bmtheta)}$ (bottom panel).
    Their expansion in the OAO basis is given in percentage, where we kept the sign of the amplitudes as it provides information on the bonding, non-bonding, or anti-bonding character of the orbital.
    Energies of the canonical orbitals are also provided in Hartree.}
    \label{fig:canonical_orbitals}
\end{figure*}

Turning to the molecular occupied orbitals, we plot in Fig.~\ref{fig:canonical_orbitals} the canonical orbitals obtained by diagonalizing the $2\times 2$ Fock matrix in the subspace spanned by the two prepared states.
For PySCF and the successful ensemble VQE noiseless simulation, this subspace is identical to the one obtained by block-diagonalizing the original $4\times 4$ Fock matrix.
Note that constructing the $2 \times 2$ occupied subspace of Fock matrix is only possible for the initial states, as well as the states from PySCF and the noiseless simulation, as the noisy simulation and the experiment do not yield pure states.
In other words, by diagonalizing the $4\times 4$ ensemble density matrices built from the aforementioned states, we obtain the eigenvalues $\{2, 2, 0, 0\}$
by construction, corresponding
to the orbital occupation.
Hence, the first two eigenstates span the occupied subspace while the last two eigenstates span the virtual (unoccupied) subspace.
However, for the noisy simulation and the experiment, we obtain $\{2.006917, 1.878224, 0.11818, -0.003322\}$ and $\{1.891369, 1.735589, 0.329119, 0.043923\}$, respectively, so that no clearly defined occupied or virtual orbitals can be extracted.

As can be seen in Fig.~\ref{fig:canonical_orbitals}, the first initial orbital is clearly a bonding orbital between the Li(1s) and H(1s) orbitals.
In contrast, the optimal occupied orbital (from the noiseless simulation) is much lower in energy (by about 0.8 Hartree) and is almost entirely localized on the lithium atom (99\%, with 96.8\% from the Li(1s) orbital and 3.1\% from the Li(2s) orbital). It is therefore a non-bonding orbital.
The second initial orbital also differs significantly from the noiseless results. Although it has strong contributions from the Li(2s) and Li(2p$_z$) orbitals, it has a small and negative contribution from the H(1s) orbital. Hence, this orbital is slightly antibonding, whereas the ideal orbital is clearly bonding between lithium and hydrogen, with contributions of 30.4\% from the Li(2s) orbital, 20.5\% from the Li(2p$_z$) orbital, and 47.7\% from the H(1s) orbital.

\begin{table*}
    \centering
    \begin{tabular}{ccccc|cccc}
        &\multicolumn{4}{c}{First occupied orbital} & \multicolumn{4}{c}{Second occupied orbital} \\
        \cmidrule{2-5} \cmidrule{6-9}
        & Li(1s) & Li(2s) & Li(2p$_z$) & H(1s) & Li(1s) & Li(2s) & Li(2p$_z$) & H(1s) \\
        \midrule
        Initial & 0.674269 & 0.212983 & 0.212983 & 0.674269 & -0.212983 & 0.674269 & 0.674269 & -0.212983 \\
        PySCF &  0.983865 & 0.175572 & -0.005355 & 0.034001 & -0.119656 & 0.550514  & 0.452890 & 0.691019 \\
        Statevector &  0.983856 & 0.175618 & -0.005403  & 0.033994  & -0.119760 & 0.551111 & 0.452373 & 0.690864 \\
        \bottomrule
    \end{tabular}
    \caption{Occupied canonical orbital coefficients in the basis of orthogonal atomic orbitals.}
\end{table*}

\end{document}